\documentclass[%
 aip,
 amsmath,amssymb,
preprint,%
]{revtex4-1}

\usepackage{graphicx}
\usepackage{dcolumn}
\usepackage{bm}

\usepackage[utf8]{inputenc}
\usepackage[T1]{fontenc}
\usepackage{mathptmx}
\usepackage{etoolbox}
\usepackage[version=4]{mhchem} 
\usepackage{siunitx} 

\usepackage[color=yellow]{todonotes}

\usepackage{hyperref}

\makeatletter
\def\@email#1#2{%
 \endgroup
 \patchcmd{\titleblock@produce}
  {\frontmatter@RRAPformat}
  {\frontmatter@RRAPformat{\produce@RRAP{*#1\href{mailto:#2}{#2}}}\frontmatter@RRAPformat}
  {}{}
}%
\makeatother
\begin{document}

\title{PINN-Based Solution for a Diffusion Controlled Droplet Growth}
\author{Pavel Gol'din}
\affiliation{Alpha SystematiX}
\author{Gennady Y. Gor}
 \email{gor@njit.edu}
 \homepage{http://porousmaterials.net}
\affiliation{Otto H. York Department of Chemical and Materials Engineering,\\
New Jersey Institute of Technology,\\
323 Dr. Martin Luther King Jr. Blvd, Newark, NJ 07102, USA}

\date{\today}

\begin{abstract}
We study diffusion-controlled growth of a spherical droplet with a moving boundary using a physics-informed neural network (PINN) formulation.
The governing diffusion equation is coupled to the interfacial mass balance, with the droplet radius treated as an additional trainable function of time.
The PINN accurately reproduces the self-similar growth law and concentration profiles for a wide range of initial droplet radii, demonstrating convergence toward the asymptotic diffusive regime.
The proposed approach provides a flexible and computationally efficient framework for solving moving-boundary diffusion problems and can be readily extended to include additional physical effects.
\end{abstract}

\maketitle

\section{Introduction}

Diffusion-driven growth of particles of a new phase is a fundamental process in various physical and chemical systems, including aerosol formation in vapor–gas mixtures, crystallization in melts or solutions. Classical theoretical descriptions of such processes are typically formulated in spherical symmetry. Even within these assumptions, solution of such problems appear challenging both analytically and numerically. The reason for that is the Stefan boundary condition which is set at the moving boundary -- the surface of the growing particle of the new phase. 

To be specific we focus on a problem of diffusion-controlled droplet grow in a vapor-gas medium~\cite{fuchs1959evaporation}, a problem central to atmospheric chemistry and  physics~\cite{seinfeld2016atmospheric}. If the initial concentration is considered uniform, initial droplet radius is taken to be vanishingly small, \( R(0) = 0 \), and the diffusion coefficient is constant, the solution of the diffusion problem can be written in the analytical self-similar form~\cite{adzhemyan2006self}, providing the simple time dependence of the droplet radius \( R(t) \propto \sqrt{t} \). This type of self-similar solution has been demonstrated earlier for crystallization in melts and solutions~\cite{zener1949theory, frank1950radially}.

Once any of the above assumptions are dropped, e.g. a finite initial radius is introduced, or the diffusion coefficient considered concentration-dependent, the self-similarity is lost, and solution of the problem presents a challenge. The analytical solution typically rely on the quasi-steady state approximation~\cite{fuchs1959evaporation, grinin2004study}. The numerical solution using finite difference technique is also challenging due to the need of adaptive mesh, etc. ~\cite{shyy1995computational}.

Physics-Informed Neural Networks (PINNs) provide an alternative and highly flexible framework for solving such problems ~\cite{lagaris1998artificial,raissi2019physics, karniadakis2021physics}. In the PINN approach, a neural network approximates the concentration governing partial differential equation, boundary conditions, and interface dynamics are imposed through the loss function. Unlike mesh-based numerical methods, PINNs can handle moving boundaries, discontinuities, and coupled nonlinear dynamics without explicit discretization. Moreover, they allow for direct optimization of additional unknowns, such as the droplet radius \( R(t) \), jointly with the concentration profile.

In this work, we apply the PINN methodology to the classical diffusion-controlled droplet growth problem originally studied by Adzhemyan et al.~\cite{adzhemyan2006self}. Without assuming self-similarity, the neural network ``learns'' both the concentration profile and the time evolution of the droplet radius. Remarkably, the trained PINN reproduces the self-similar scaling \( R(t) \propto \sqrt{t} \) and yields concentration profiles consistent with the analytical solution, even when the initial radius is finite. This demonstrates that PINNs not only recover known analytical results as emergent features, but also extend them to regimes where analytical methods fail, providing a unified, data-driven framework for studying diffusion processes with moving boundaries. 

\section{Theoretical Background: Diffusion-Controlled Droplet Growth}

The growth of a liquid droplet in a supersaturated vapor–gas mixture is a classical diffusion-controlled process. In the diffusion-limited regime, the transport of mass from the vapor to the droplet surface is much slower than the interfacial kinetics of condensation. Therefore, the surrounding medium can be described by the diffusion equation in spherical symmetry,
\begin{equation}
    \frac{\partial n}{\partial t} 
    = D \left( 
    \frac{\partial^2 n}{\partial r^2} 
    + \frac{2}{r}\frac{\partial n}{\partial r} 
    \right),
    \label{eq:diffusion}
\end{equation}
with boundary conditions
\begin{equation}
    n(R(t),t) = n_{\rm s}, 
    \qquad 
    \lim_{r \to \infty}n(r,t) = n_0,
    \label{eq:boundary}
\end{equation}
where \(R(t)\) is the droplet radius, \(D\) is the diffusion coefficient, and \(n_{\rm s}\) and \(n_0\) are the vapor concentrations (number densities) at the droplet surface and in the bulk phase, respectively. The motion of the liquid–vapor interface is governed by conservation of mass at the surface (Stefan condition),
\begin{equation}
    \left( n_{\rm l} - n_0 \right) \frac{{\rm d}R}{{\rm d}t}
    = 
    D \left. \frac{\partial n}{\partial r} \right|_{r = R(t)},
    \label{eq:stefan}
\end{equation}
where \( n_{\rm l} \) is the number density of the liquid phase.

We introduce the dimensionless difference in concentration 
\begin{equation}
\label{eq:phi}
    \phi(r,t) \equiv \frac{n(r,t) - n_{\rm s}}{n_0 - n_{\rm s}}.
\end{equation}
We also introduce some characteristic length scale $\ell$, and corresponding time scale $t_0 \equiv \ell^2/D$. We will use the following dimensionless variables to rewrite the problem:
\begin{equation}
\label{eq:dimensionless}
    \tilde{R} \equiv R/\ell \qquad \tilde{r} \equiv r/\ell \qquad \tilde{t} \equiv t/t_0.
\end{equation}
Using Eqs.~\ref{eq:phi} and \ref{eq:dimensionless} in Eq.~\ref{eq:diffusion} we get
\begin{equation}
    \frac{\partial \phi}{\partial \tilde{t}} 
    = \left( 
    \frac{\partial^2 \phi}{\partial \tilde{r}^2} 
    + \frac{2}{\tilde{r}}\frac{\partial \phi}{\partial \tilde{r}} 
    \right),
    \label{eq:diffusion1}
\end{equation}
and the boundary conditions read
\begin{equation}
    \phi(\tilde{R}(\tilde{t}),\tilde{t}) = 0, 
    \qquad 
    \lim_{\tilde{t} \to \infty}\phi(\tilde{r},\tilde{t}) = 1,
    \label{eq:boundary1}
\end{equation}
and 
\begin{equation}
    \frac{{\rm d}\tilde{R}}{{\rm d}\tilde{t}} = a \left. \frac{\partial \phi}{\partial \tilde{r}} \right|_{\tilde{r} = \tilde{R}(\tilde{t})},
    \label{eq:stefan1}
\end{equation}
where
\begin{equation}
    \label{eq:a}
    a \equiv \frac{n_0 - n_{\rm s}}{n_{\rm l} - n_0}.
\end{equation}

Following Adzhemyan et al.~\cite{adzhemyan2006self}, one can introduce a dimensionless variable 
\begin{equation}
    \rho \equiv \frac{r}{R(t)} = \frac{\tilde{r}}{\tilde{R}(\tilde{t})},
    \label{eq:ansatz}
\end{equation}
the radial coordinate normalized by the instantaneous droplet radius, so that \( \rho = 1 \) corresponds to the moving interface. Substituting Eq.~\eqref{eq:ansatz} into Eq.~\eqref{eq:diffusion1} leads to the ordinary differential equation
\begin{equation}
    \phi''(\rho) + \frac{2}{\rho} \phi'(\rho) + b\,\rho \phi'(\rho) = 0,
    \label{eq:automodel}
\end{equation}
where \( b \) is a constant to be determined. The corresponding growth law for the droplet radius follows from the requirement that the similarity form \eqref{eq:ansatz} remains valid for all \(t\):
\begin{equation}
    \tilde{R}^2(\tilde{t}) = 2b \tilde{t}.
    \label{eq:growth}
\end{equation}
Thus, the droplet radius scales as \( R(t) \propto \sqrt{t} \), which is a direct consequence of diffusion-limited transport. 

The self-similar solution of this problem reads~\cite{zener1949theory, frank1950radially, adzhemyan2006self}:
\begin{equation}
    \label{eq:phi-solution}
    \phi(\rho) = \frac{b}{a} \int\limits_{1}^{\rho}
    \frac{{\rm d}x}{x^2} \exp\!\left[-\frac{b}{2}(x^2 - 1)\right] \, 
\end{equation}
where the parameter \(b\) is uniquely determined from the Stefan condition \eqref{eq:stefan}. Substituting the self-similar form (Eq.~\ref{eq:phi-solution}) into the Stefan condition yields the following implicit relation for the parameter \( b \):
\begin{equation}
    a = b \int\limits_{1}^{\infty}
    \frac{{\rm d}x}{x^2} \exp\!\left[-\frac{b}{2}(x^2 - 1)\right] \,.
    \label{eq:b_condition}
\end{equation}

Although this self-similar solution is elegant and physically transparent, it relies on several restrictive assumptions:  (i) the initial droplet radius is vanishingly small, \( R(0) = 0 \);  (ii) the initial concentration is uniform; (iii) the diffusion coefficient \(D\) is constant. In real systems, however, the initial radius is finite, early-time transients are not self-similar, and transport coefficients may depend on temperature or concentration~\cite{fuchs1959evaporation}. In such regimes, no analytical solution exists. In the following sections, we demonstrate that a Physics-Informed Neural Network (PINN) can recover the self-similar solution (Eq.~\ref{eq:phi-solution}) without imposing it explicitly, and can also extend the analysis to cases where self-similarity does not strictly apply, such as finite initial radius.

\section{Physics-Informed Neural Networks}

Physics-Informed Neural Networks (PINNs) have emerged as a general framework for solving partial differential equations (PDEs) by embedding the governing physical laws into the structure of a neural network~\cite{lagaris1998artificial, raissi2019physics, karniadakis2021physics}. In their standard formulation, a neural network \( \phi(\mathbf{x}, t; \theta) \), parameterized by trainable weights \( \theta \), approximates the solution of a PDE. The network receives spatial and temporal coordinates as inputs and returns the corresponding profile value (e.g., concentration, temperature, or displacement). Instead of minimizing a loss defined only by data, the PINN loss function incorporates the residual of the governing equation itself, ensuring that the predicted solution satisfies the physics everywhere in the domain.

For a generic PDE of the form
\[
\mathcal{N}[\phi(\mathbf{x},t)] = 0,
\]
the residual \( \mathcal{R}(\mathbf{x},t) = \mathcal{N}[\phi(\mathbf{x},t)] \) is computed by automatic differentiation and included in the total loss function:
\[
\mathcal{L} = \lambda_{\text{PDE}}\langle |\mathcal{R}(\mathbf{x},t)|^2\rangle + \lambda_{\text{BC}}\langle |\phi - \phi_{\text{BC}}|^2\rangle + \lambda_{\text{IC}}\langle |\phi - \phi_{\text{IC}}|^2\rangle.
\]
The terms correspond to the PDE residual, boundary conditions (BC), and initial condition (IC), respectively, and \( \lambda_i \) are weighting coefficients. The minimization of this loss function drives the network toward a solution that satisfies both the physical constraints and any available data. This approach eliminates the need for spatial or temporal discretization and has been successfully applied to a wide range of steady and transient problems, including heat conduction~\cite{cai2021physics}, wave propagation~\cite{song2022versatile, rasht2022physics}, and fluid dynamics~\cite{steinfurth2024assimilating, rudenko2025reconstruction}.

Despite its flexibility, the standard PINN formulation assumes that the computational domain and its boundaries are fixed and known a priori. Consequently, PINNs perform well for problems with stationary boundaries or prescribed boundary motion but face significant challenges when the domain itself evolves in time. Diffusion-controlled phase change, such as the growth of a droplet in a supersaturated vapor, falls precisely into this latter category: the moving boundary between the liquid and vapor phases, \( r = R(t) \), must be determined simultaneously with the concentration profile. The dynamics of this interface depends on the gradient of concentration at the droplet surface, introducing a nonlinear coupling between the PDE and an additional ordinary differential equation (ODE) governing \( R(t) \).

In such cases, the classical PINN approach becomes inadequate because it cannot directly represent or update the unknown boundary position during training. To overcome this limitation, we extend the standard PINN formulation by treating the droplet radius \( R(t) \) as a trainable function represented by a set of learnable parameters. The evolution of \( R(t) \) is constrained through an additional loss term derived from the interfacial mass balance condition. This modification allows the network to self-consistently learn both the diffusion profile and the moving boundary dynamics within a single optimization framework. As shown below, this generalized PINN not only reproduces the analytical self-similar solution by  Adzhemyan et al. but also remains valid for finite initial radii and transient non-self-similar regimes.

\subsection{Implementation Details}

To solve the diffusion-controlled droplet growth problem with a moving boundary, we adopt a modified Physics-Informed Neural Network (PINN) formulation capable of simultaneously learning the diffusion profile and the evolution of the droplet radius. While the numerical solution uses dimensionless variables $\tilde{R}$, $\tilde{r}$, and $\tilde{t}$, we are going to omit the tildes for shortness. The dimensionless concentration profile \( \phi(r,t) \) is represented by a fully connected feedforward neural network (denoted as \texttt{PhiNet}), which takes as inputs the radial coordinate \(r\) and time \(t\), and outputs the corresponding value of \(\phi\). The network consists of four hidden layers with 15 neurons per layer and uses the hyperbolic tangent (\texttt{tanh}) activation function, which provides smoothness and differentiability required for automatic differentiation.

Unlike a conventional PINN, where all domain boundaries are fixed, the present formulation introduces an additional set of trainable parameters that represent the droplet radius \( R(t) \) at discrete time nodes. These parameters, denoted as \( R_i \equiv R(t_i) \), are optimized jointly with the neural network weights. During training, the radius at intermediate times is obtained by linear interpolation between the neighboring time nodes, ensuring that the interface position evolves continuously in time. This treatment effectively embeds the moving boundary into the learning process, eliminating the need for explicit front tracking or re-meshing.

The loss function combines several physically motivated terms:
\begin{equation}
\mathcal{L} = 
\mathcal{L}_{\text{PDE}} +
\mathcal{L}_{\text{IC}} +
\mathcal{L}_{\text{BC}} +
\mathcal{L}_{R(t)} +
\mathcal{L}_{R_0}.
\label{eq:loss_total}
\end{equation}
Here, \( \mathcal{L}_{\text{PDE}} \) enforces the diffusion equation in spherical coordinates, Eq.~\ref{eq:diffusion1},
\[
\mathcal{L}_{\text{PDE}} = \left\langle
\left(
\frac{\partial \phi}{\partial t}
- \frac{\partial^2 \phi}{\partial r^2}
- \frac{2}{r}\frac{\partial \phi}{\partial r}
\right)^2
\right\rangle,
\]
where the angle brackets denote averaging over collocation points \((r,t)\) in the spatio-temporal domain. The terms \( \mathcal{L}_{\text{IC}} \) and \( \mathcal{L}_{\text{BC}} \) correspond to the initial and boundary conditions, respectively. The condition at the moving interface is expressed as \(\phi(R(t),t)=0\), while the initial concentration profile satisfies \(\phi(r,0)=1\).

A crucial component of this formulation is the term \( \mathcal{L}_{R(t)} \), which constrains the droplet radius dynamics according to the interfacial flux condition, Eq.~\ref{eq:stefan1}. In practice, the derivative \(\partial \phi / \partial r\) is evaluated at the instantaneous interface position \(r = R_i\), and the temporal derivative of \(R(t)\) is approximated by a finite difference between successive time nodes. This coupling between the PDE residual and the interface condition ensures that the growth rate of the droplet is consistent with the diffusive flux at its surface.

The final term, \( \mathcal{L}_{R_0} = (R(0) - R_0)^2 \), enforces the prescribed initial radius. The total loss \(\mathcal{L}\) is minimized with respect to both the neural network weights and the radius parameters using the Adam optimizer. During training, automatic differentiation is employed to compute spatial and temporal derivatives of \(\phi(r,t)\), enabling accurate enforcement of the diffusion equation without explicit discretization.

This formulation allows the neural network to ``discover'' both the concentration profile and the time evolution of the droplet radius self-consistently. Remarkably, the trained PINN reproduces the self-similar scaling \( R(t) \propto \sqrt{t} \) predicted by the analytical theory of Adzhemyan et al., even though self-similarity is not imposed explicitly. Furthermore, the method remains robust for finite initial radii and non-self-similar transients, demonstrating its ability to generalize the classical diffusion-controlled growth model beyond the assumptions of strict self-similarity.

\subsubsection{Training Algorithm}

The learning procedure for the proposed PINN can be summarized as follows:

\begin{enumerate}
    \item Initialize the neural network parameters \( \theta \) for the concentration profile \( \phi(r,t;\theta) \).
    \item Define the discrete set of trainable radius values \( R_i = R(t_i) \) at temporal nodes \( t_i \in [0,1] \).
    \item For each training epoch:
    \begin{enumerate}
        \item Generate collocation points \( (r,t) \) in the domain \( r \in [R(t), R_{\text{max}}] \).
        \item Evaluate the PDE residual of the diffusion equation:
        \[
        \mathcal{R}_{\text{PDE}} = 
        \frac{\partial \phi}{\partial t}
        - \frac{\partial^2 \phi}{\partial r^2}
        - \frac{2}{r}\frac{\partial \phi}{\partial r}.
        \]
        \item Compute the losses corresponding to:
        \begin{itemize}
            \item the PDE residual (\(\mathcal{L}_{\text{PDE}}\)),
            \item the boundary and initial conditions (\(\mathcal{L}_{\text{BC}}\), \(\mathcal{L}_{\text{IC}}\)),
            \item the radius dynamics:
            \[
            \mathcal{L}_{R(t)} =
            \left\langle 
            \left( 
            \frac{{\rm d}R}{{\rm d}t} - 
            a \frac{\partial \phi}{\partial r}\big|_{r=R(t)}
            \right)^2 
            \right\rangle,
            \]
            \item and the initial radius constraint (\(\mathcal{L}_{R_0}\)).
        \end{itemize}
        \item Form the total loss:
        \[
        \mathcal{L} = \mathcal{L}_{\text{PDE}} + \mathcal{L}_{\text{IC}} + 
        \mathcal{L}_{\text{BC}} + \mathcal{L}_{R(t)} + \mathcal{L}_{R_0}.
        \]
        \item Update \( \theta \) and all \( R_i \) simultaneously by gradient descent using the Adam optimizer.
    \end{enumerate}
    \item Continue the training until the total loss \(\mathcal{L}\) falls below \(10^{-3}\).
\end{enumerate}

After convergence, the network yields the concentration profile \(\phi(r,t)\) and the droplet radius \(R(t)\) consistent with the diffusion equation and the interfacial flux condition. The learned radius evolution exhibits the self-similar scaling \( R(t) \propto \sqrt{t} \), confirming that the PINN has self-consistently recovered the Adzhemyan et al. solution without any explicit assumption of self-similarity.

\subsubsection{Training Setup}

All computations were performed using the \texttt{PyTorch} framework with automatic differentiation enabled for both spatial and temporal variables. In each experiment, the concentration profile \(\phi(r,t)\) was represented by a fully connected neural network, and the droplet radius \(R(t)\) was parameterized as a set of learnable values \(R_i = R(t_i)\) defined on a uniform temporal grid. The network parameters and the radius values were optimized jointly using the Adam optimizer with a learning rate of \(10^{-4}\).

The spatial domain was truncated at $R_{\max} = 17$, which is sufficiently large for the concentration profile to approach its far-profile value; thus, the domain effectively represents an unbounded medium. The temporal interval $[0,1]$ was discretized into $N_t = 25$ uniformly spaced nodes, corresponding to a set of learnable radius values $R(t_i)$. This number of nodes provides enough flexibility to resolve the time evolution of the interface while keeping the number of trainable parameters moderate and the optimization stable. 

At each training epoch, $1024$ collocation points $(r,t)$ were sampled to enforce the diffusion equation. This sampling density was found to provide reliable control of the PDE residual without introducing significant computational overhead. Additional point sets were generated to impose the initial condition $\phi(r,0)=1$ and the interfacial boundary condition $\phi(R(t),t)=0$. The interfacial mass balance condition linking $\dfrac{{\rm d}R}{{\rm d}t}$ to the concentration gradient at $r=R(t)$ (Eq.~\ref{eq:stefan1}) was incorporated as a separate loss term, enabling the network to learn the interface motion self-consistently. Training was performed for several different initial radii $R_0$ to investigate how the transient growth behavior depends on the initial droplet size. Each training run consisted of up to $7.5\times10^{4}$ epochs, although convergence was typically achieved earlier, once the total loss dropped below $2\times10^{-3}$. 
All computations were carried out on a standard desktop CPU, demonstrating that the proposed PINN formulation remains computationally efficient even without GPU acceleration.

\section{Results and Discussion}

\begin{figure}[t]
  \centering
  \includegraphics[width=0.85\linewidth]{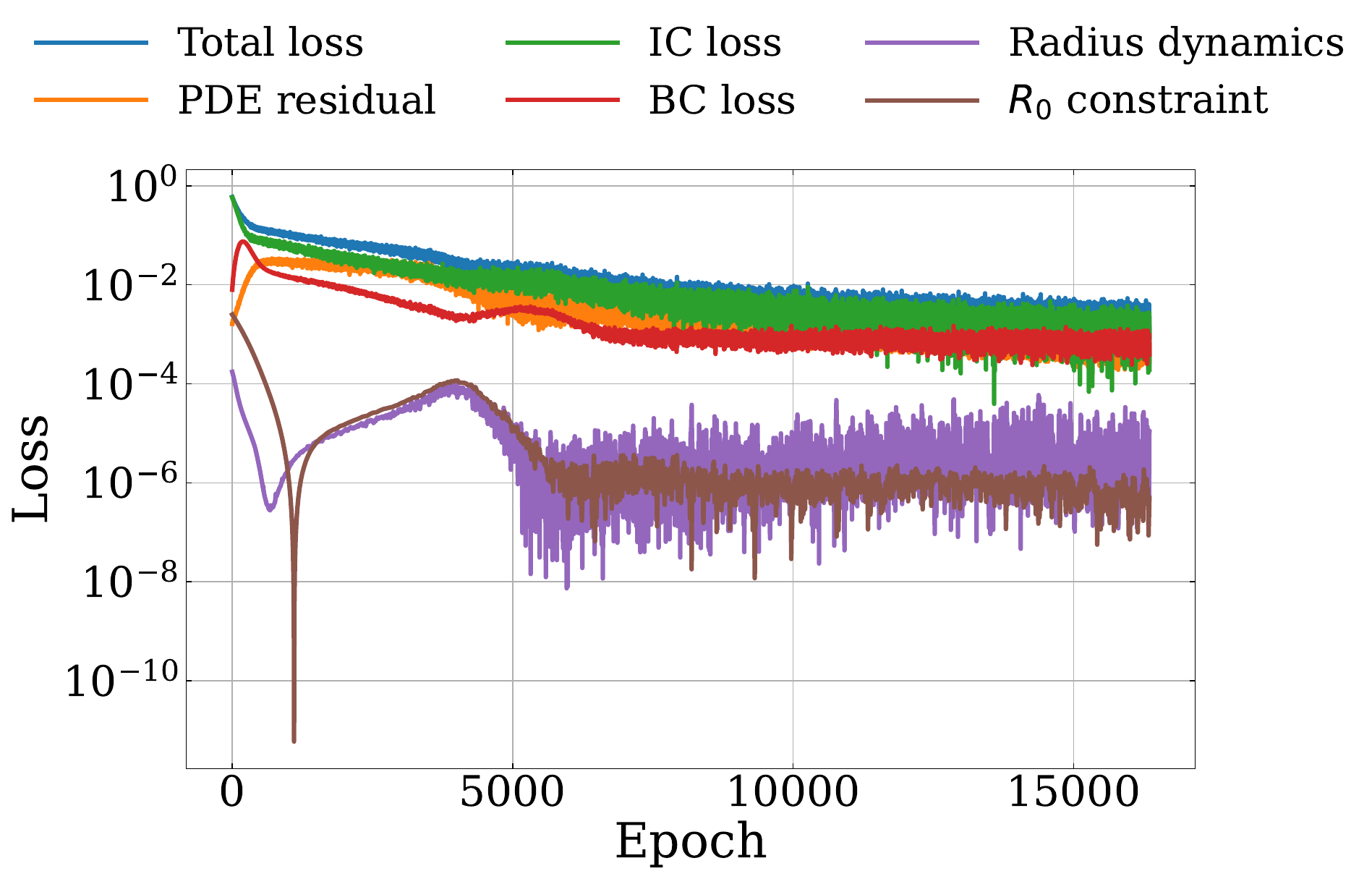}
  \caption{
    Training loss components during PINN optimization.
    The PDE residual, initial- and boundary-condition losses, and the 
    radius–dynamics constraint all decrease monotonically over the course 
    of training. The total loss stabilizes once all components reach the 
    $10^{-3}$--$10^{-5}$ level, indicating that the PINN has simultaneously 
    satisfied the diffusion equation, interface conditions, and the 
    evolution law for the droplet radius.
  }
  \label{fig:loss_components}
\end{figure}

\begin{figure}[htbp]
    \centering
    \includegraphics[width=0.70\textwidth]{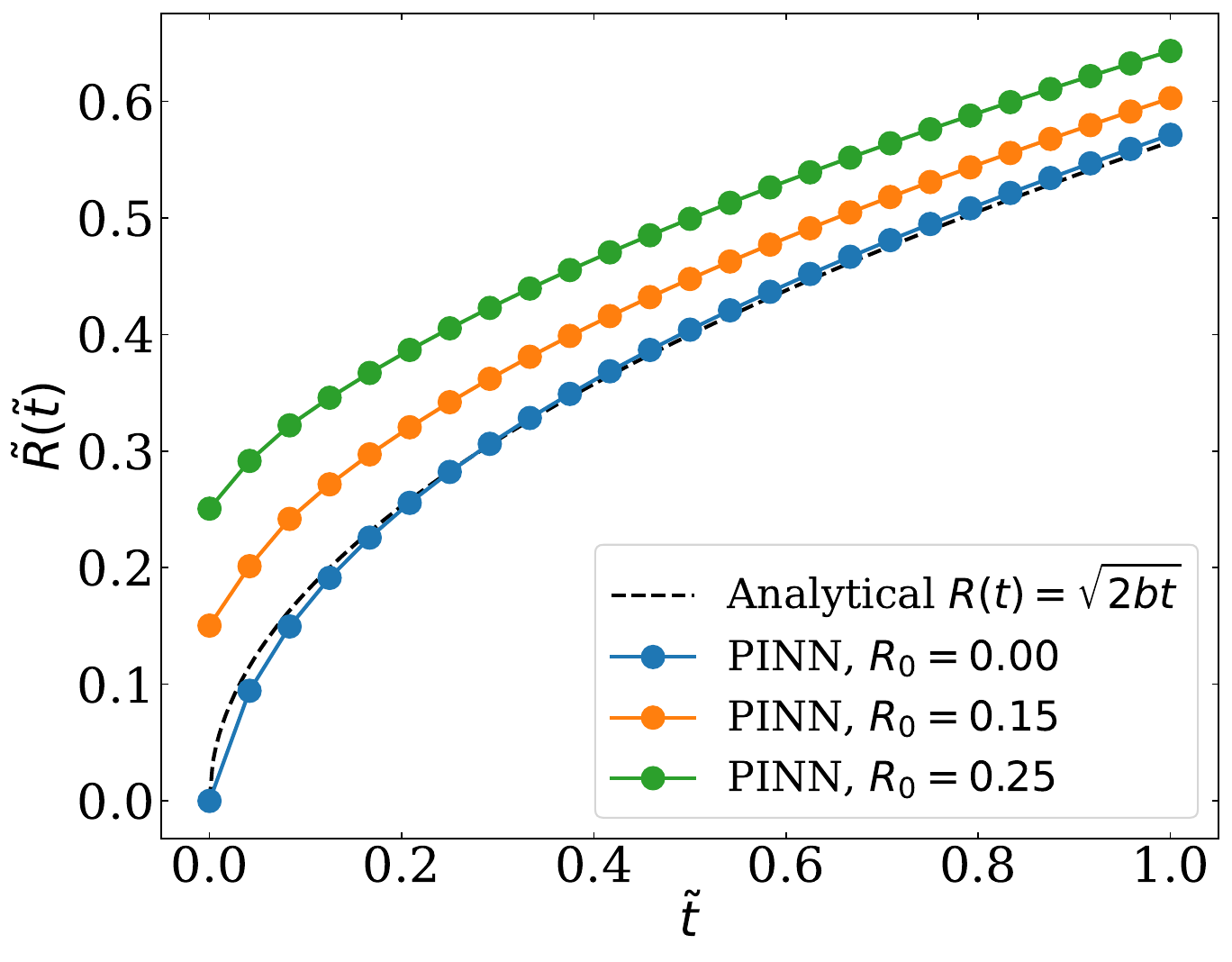}
    \caption{PINN predictions for the evolution of the droplet radius for several initial radii $R_0$. Model parameters: $a = 0.1$; the corresponding value $b = 0.159$. }
    \label{fig:R_multi}
\end{figure}

\begin{figure}[htbp]
    \centering
    \includegraphics[width=0.70\textwidth]{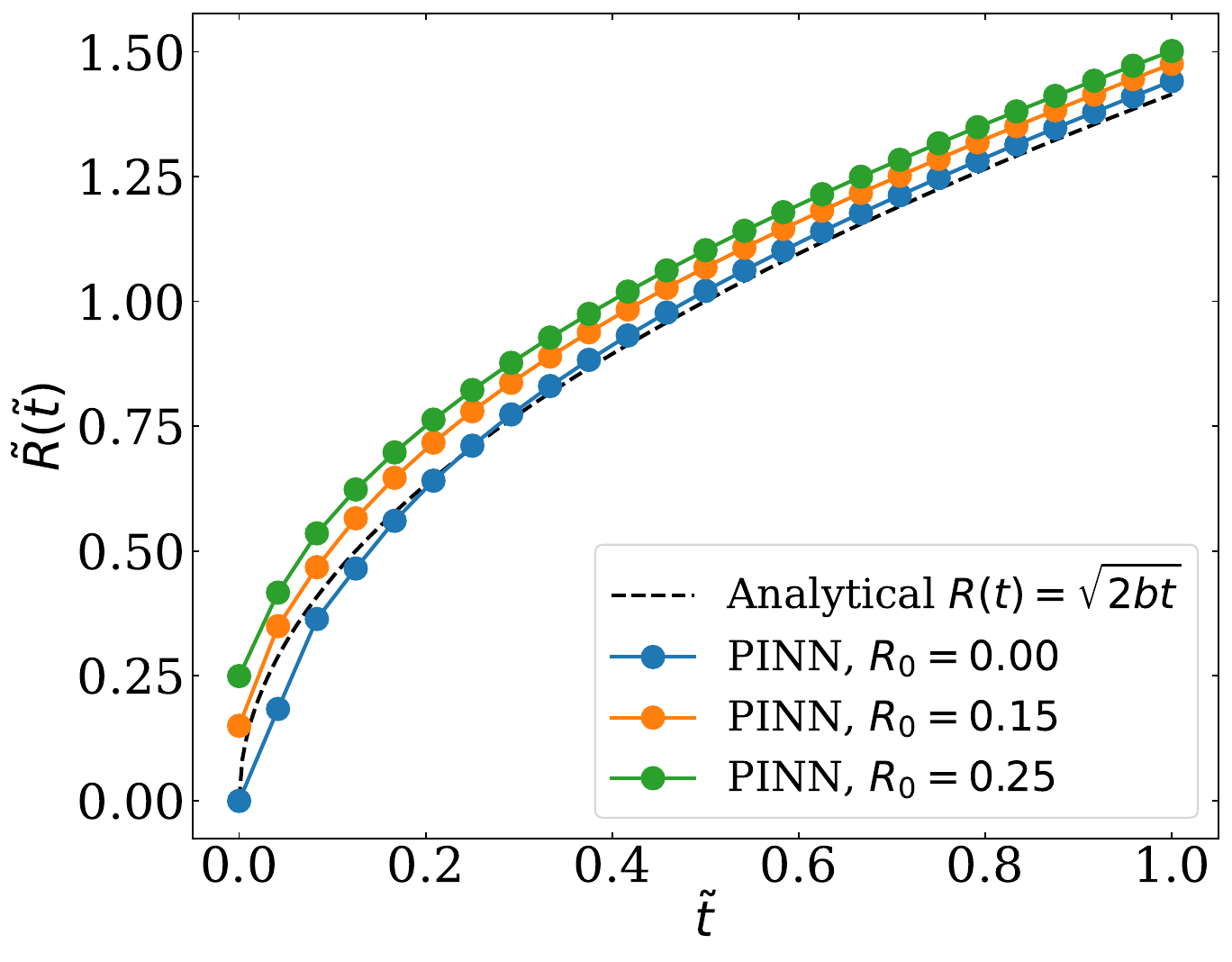}
    \caption{
     For the normalized case $b=1$, corresponding to $a=0.3445$, the PINN solution
still reproduces the overall diffusive growth trend, but the agreement with the
analytical profiles is slightly reduced compared to the slower-growth
case $b=0.159$. This behavior is expected, as larger values of $b$ correspond to
faster interface motion and stronger gradients in the concentration profile,
which make the learning problem more challenging.
    }
    \label{fig:R_multi_big}
\end{figure}

\begin{figure}[htbp]
  \centering
  \includegraphics[width=\linewidth]{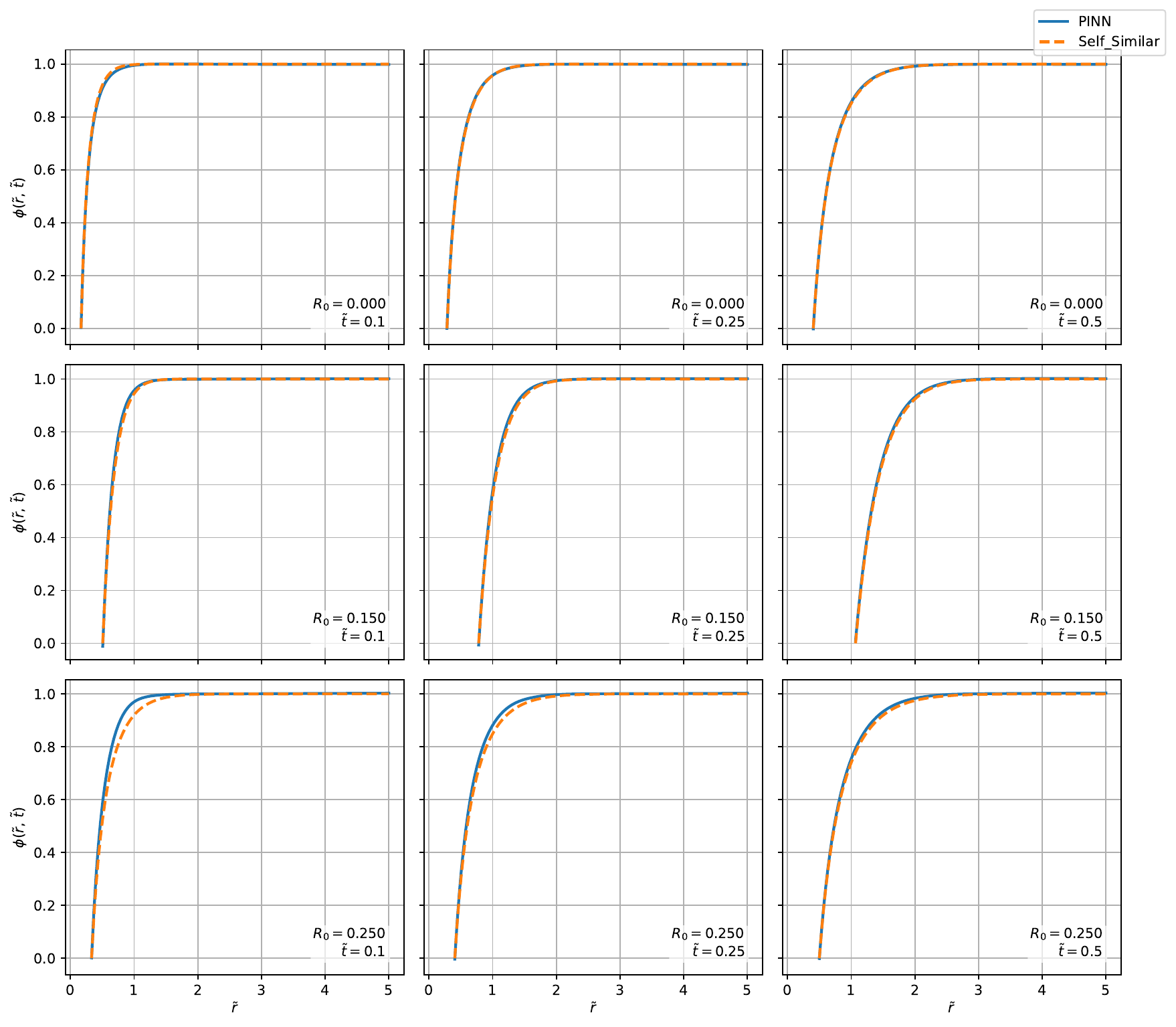}
  \caption{
    PINN and self-similar concentration profiles for different initial radii $R_0$ and reduced times $\tilde t$. Each panel shows a pair of curves: the solid line corresponds to the PINN solution, while the dashed line represents the analytical self-similar profile.
  }
  \label{fig:profiles_R0xT}
\end{figure}

Figure~\ref{fig:loss_components} shows the evolution of all loss components during training. The PDE residual and the boundary and initial condition terms exhibit steady decay over several orders of magnitude, while the auxiliary loss enforcing the interfacial radius dynamics follows a similar trend. After approximately $2\times 10^{3}$ epochs, all loss terms fall below $10^{-3}$, and the total loss plateaus, indicating that the model has simultaneously satisfied the diffusion equation, the interface conditions, and the learned trajectory of the droplet radius. All loss components were assigned equal weights. For the considered problem, the optimization demonstrated stable and simultaneous convergence of all constraints, indicating that additional loss reweighting was not required. This stabilization correlates with the emergence of the correct self-similar growth law observed in the learned radius and concentration profiles shown in Figures~\ref{fig:R_multi}, \ref{fig:R_multi_big}, \ref{fig:profiles_R0xT}.

Figure~\ref{fig:R_multi} presents the evolution of the droplet radius for several different initial radii $R_0$. Although the trajectories start from distinct initial sizes, the subsequent growth rapidly approaches a common diffusive regime. In all cases, the numerical solution converges toward the analytical prediction $R(t) = \sqrt{2 b t}$, which corresponds to the self-similar limit obtained for a vanishing initial radius (Eq.~\ref{eq:growth}).

This behavior reflects the fact that the term $R_0^2$ influences only the short-time dynamics: it introduces a horizontal shift in the early evolution of $R^2(t)$ but does not alter the long-time growth rate. Once the transient stage is passed, the interface motion becomes fully diffusion-controlled, and the solution enters the universal self-similar regime. The excellent
agreement between the PINN trajectories and the analytical curve for all tested $R_0$ demonstrates that the network correctly captures both the diffusion profile and the moving-boundary kinetics, recovering the expected self-similar growth law without explicitly imposing it. 

The multi-panel comparison in Fig.~\ref{fig:profiles_R0xT} illustrates how the PINN-predicted concentration profile approaches the self-similar profile for different initial droplet radii $R_0$ and reduced times $\tilde t$. For each $(R_0,\tilde t)$ pair, the PINN solution (solid line) and the analytical reference curve (dashed line) exhibit excellent agreement across the entire outer region $\tilde r > \tilde R(t)$.
At early times, when the influence of the finite initial radius is still present, small deviations between the two profiles are visible, primarily in the vicinity of the moving interface. However, these differences rapidly diminish as $\tilde t$ increases. For all three initial radii, the profiles progressively collapse onto the same universal shape, confirming the emergence of the self-similar regime.
Overall, the figure demonstrates that the PINN accurately reconstructs both the spatial structure of the concentration profile and its convergence toward the asymptotic self-similar form, irrespective of the initial droplet size.

The PINN accurately reproduces the droplet growth dynamics for all tested initial radii $R_0$.
As shown in Fig.~\ref{fig:R_multi}, the learned radius progressively approaches the asymptotic scaling $R(t)\sim\sqrt{2 b t}$ as the influence of the initial radius diminishes. The agreement between the PINN trajectories and the analytical prediction is excellent across the entire time interval.

Figure~\ref{fig:profiles_R0xT} compares the concentration profiles produced by the PINN with the self-similar solution. At early times, small deviations appear near the interface due to the finite initial radius, but these rapidly vanish. For all $R_0$ and for all tested times, the PINN captures the correct spatial structure and converges toward the universal self-similar profile.
Overall, the results demonstrate that the proposed PINN formulation reliably recovers both the
correct growth law and the correct concentration profile, even when the initial droplet radius is
finite.

\begin{figure}[t]
  \centering
  \includegraphics[width=0.8\linewidth]{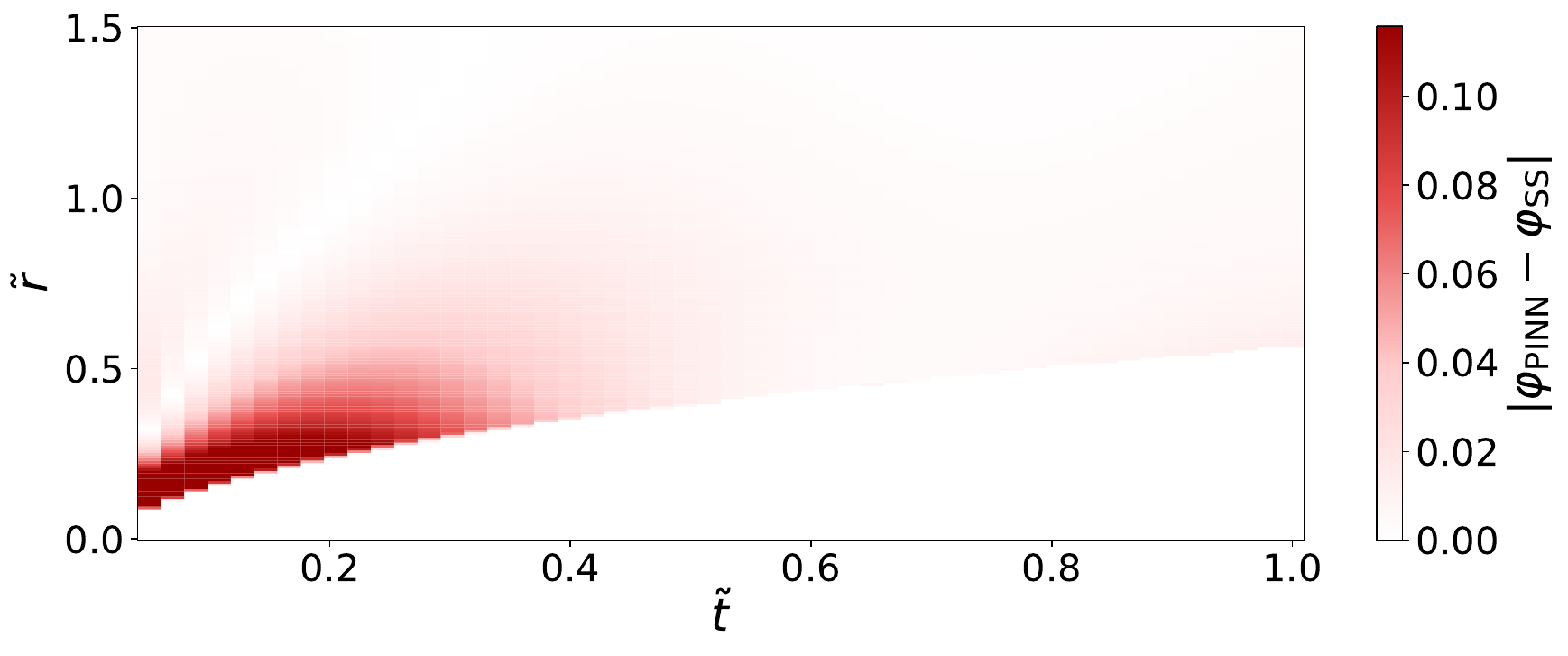}
  \caption{
    Space--time map of the absolute error between the PINN prediction
    and the self-similar solution, $|\varphi_{\mathrm{PINN}} - \varphi_{\mathrm{SS}}|$,
    for $R_0 = 0$. The error is localized near the moving interface at early times
    and rapidly decays both in space and in time, indicating that the
    PINN accurately reproduces the concentration profile over most of
    the domain.
  }
  \label{fig:error_heatmap}
\end{figure}

\begin{figure}[t]
  \centering
  \includegraphics[width=0.8\linewidth]{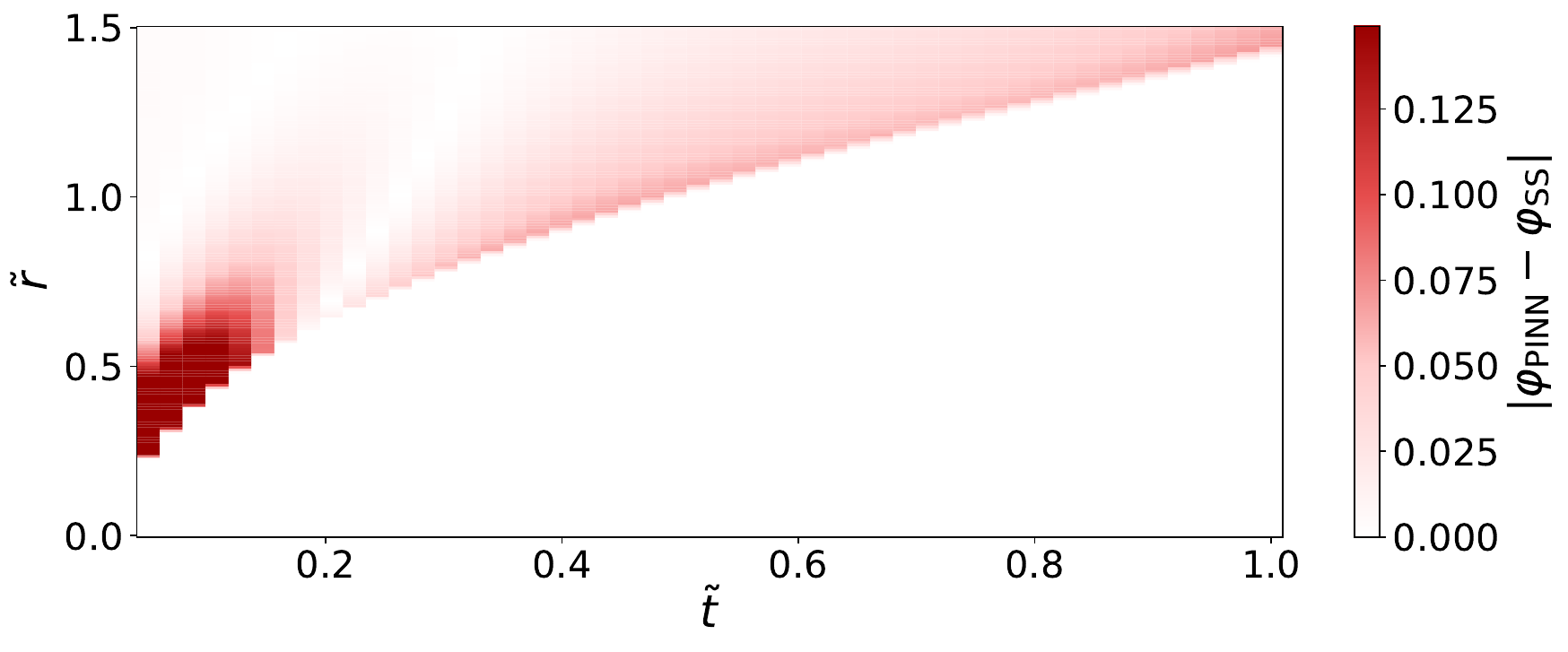}
  \caption{
   Despite the higher growth rate corresponding to $b=1$, the PINN remains stable and accurately reproduces the qualitative structure of the concentration profile,
as evidenced by the error heat map.  }
  \label{fig:error_heatmap2}
\end{figure}

To quantify the accuracy of the learned concentration profile in the
entire space--time domain, we computed the absolute deviation between
the PINN prediction and the self-similar solution, $|\varphi_{\mathrm{PINN}} - \varphi_{\mathrm{SS}}|$.
The resulting error map is shown in Fig.~\ref{fig:error_heatmap}.
The error is mainly concentrated in a narrow region close by to the moving droplet interface at early times, where the concentration gradient is the largest, and quickly decays away from the front, confirming that the PINN reconstructs the self-similar concentration profile with high fidelity.

An important advantage of the PINN framework employed here is its flexibility with respect to modifications of the governing equations and boundary conditions. In particular, the present formulation can be naturally extended to account for additional physical effects at the moving interface, such as curvature-dependent boundary conditions arising from surface tension (Kelvin effect)~\cite{fuchs1959evaporation,seinfeld2016atmospheric}. In this case, the equilibrium concentration at the droplet surface becomes a function of the local curvature, leading to a nonlinear coupling between the interface geometry and the diffusion profile ~\cite{grinin2004study,adzhemyan2006self}. Incorporating such effects within traditional numerical schemes typically requires nontrivial interface tracking and mesh adaptation  ~\cite{shyy1995computational}, whereas within the PINN approach they can be included directly through additional terms in the loss function.

\section{Conclusion}

In this work, we have demonstrated that a physics-informed neural network can successfully reproduce the classical diffusion-controlled growth of a spherical droplet with a moving boundary. Without imposing any self-similar structure \emph{a priori}, the PINN formulation recovers both the diffusive growth law for the droplet radius and the spatial structure of
the concentration profile predicted by the analytical solution. 
 The concentration profiles obtained from the PINN show excellent  agreement with the analytical self-similar solution over a wide range of times, while the error heat maps indicate that discrepancies are primarily confined to early transient stages and the vicinity of the moving interface.  
We have also examined a faster growth regime corresponding to a larger value of the dimensionless parameter $b$. While the self-similar behavior is still recovered, both the radius and concentration profiles exhibit slightly larger deviations during the transient stage, reflecting the increased stiffness of the problem. Nevertheless, the overall agreement remains robust, demonstrating the stability of the proposed approach across different growth rates.  The results highlight the flexibility of PINNs for moving-boundary diffusion problems and their ability to capture emergent asymptotic regimes directly from the governing equations and boundary conditions. The present framework can be naturally extended to more complex situations, including curvature-dependent interfacial conditions, non-isothermal effects, or coupled transport processes. These directions open the way for applying physics-informed machine learning to a broad class of classical and modern problems describing growth of particles of new phase and evolving interfaces.

\section*{Data Availability}

The code used to generate the results presented in this paper is publicly available at \href{https://github.com/PavelGoldin1981/PINN_Moving_Boundary}{GitHub}.

\section*{Acknowledgments}

The authors thank Dr. Ilya Kremnev for inspiring the collaboration.  

\newpage

\bibliography{pinn,drop}

@book{seinfeld2016atmospheric,
  title={Atmospheric chemistry and physics: from air pollution to climate change},
  author={Seinfeld, John H and Pandis, Spyros N},
  year={2016},
  publisher={John Wiley \& Sons}
}

@book{fuchs1959evaporation,
  title={Evaporation and droplet growth in gaseous media},
  author={Fuchs, Nikolai Albertovich},
  year={1959},
  publisher={Pergamon Press}
}

@article{grinin2004study,
  title={Study of nonsteady diffusional growth of a droplet in a supersaturated vapor: Treatment of the moving boundary and material balance},
  author={Grinin, A. P. and Shchekin, A. K. and Kuni, F. M. and Grinina, E. A. and Reiss, H.},
  journal={The Journal of Chemical Physics},
  volume={121},
  number={1},
  pages={387--393},
  year={2004},
  publisher={American Institute of Physics}
}

@article{frank1950radially,
  title={Radially symmetric phase growth controlled by diffusion},
  author={Frank, Frederick Charles},
  journal={Proceedings of the Royal Society of London. Series A. Mathematical and Physical Sciences},
  volume={201},
  number={1067},
  pages={586--599},
  year={1950},
  publisher={The Royal Society London}
}

@article{adzhemyan2006self,
  title={Self-similar solution to the problem of vapor diffusion toward the droplet nucleated and growing in a vapor-gas medium},
  author={Adzhemyan, L. T. and Vasil’ev, A. N. and Grinin, A. P. and Kazansky, A. K.},
  journal={Colloid Journal},
  volume={68},
  pages={381--383},
  year={2006},
  publisher={Springer}
}

@article{zener1949theory,
  title={Theory of growth of spherical precipitates from solid solution},
  author={Zener, Clarence},
  journal={Journal of applied physics},
  volume={20},
  number={10},
  pages={950--953},
  year={1949},
  publisher={American Institute of Physics}
}

@book{shyy1995computational,
  title={Computational fluid dynamics with moving boundaries},
  author={Shyy, Wei},
  year={1995},
  publisher={CRC Press}
}

@article{lagaris1998artificial,
  title={Artificial neural networks for solving ordinary and partial differential equations},
  author={Lagaris, Isaac E and Likas, Aristidis and Fotiadis, Dimitrios I},
  journal={IEEE Transactions on Neural Networks},
  volume={9},
  number={5},
  pages={987--1000},
  year={1998},
  publisher={IEEE}
}

@article{raissi2019physics,
  title={Physics-informed neural networks: A deep learning framework for solving forward and inverse problems involving nonlinear partial differential equations},
  author={Raissi, Maziar and Perdikaris, Paris and Karniadakis, George E},
  journal={Journal of Computational Physics},
  volume={378},
  pages={686--707},
  year={2019},
  publisher={Elsevier}
}

@article{karniadakis2021physics,
  title={Physics-informed machine learning},
  author={Karniadakis, George Em and Kevrekidis, Ioannis G and Lu, Lu and Perdikaris, Paris and Wang, Sifan and Yang, Liu},
  journal={Nature Reviews Physics},
  volume={3},
  number={6},
  pages={422--440},
  year={2021},
  publisher={Nature Publishing Group}
}

@article{cai2021physics,
  title={Physics-informed neural networks for heat transfer problems},
  author={Cai, Shengze and Wang, Zhicheng and Wang, Sifan and Perdikaris, Paris and Karniadakis, George Em},
  journal={Journal of Heat Transfer},
  volume={143},
  number={6},
  pages={060801},
  year={2021},
  publisher={American Society of Mechanical Engineers}
}

@article{song2022versatile,
  title={A versatile framework to solve the Helmholtz equation using physics-informed neural networks},
  author={Song, Chao and Alkhalifah, Tariq and Waheed, Umair Bin},
  journal={Geophysical Journal International},
  volume={228},
  number={3},
  pages={1750--1762},
  year={2022},
  publisher={Oxford University Press}
}

@article{rasht2022physics,
  title={{Physics-informed neural networks (PINNs) for wave propagation and full waveform inversions}},
  author={Rasht-Behesht, Majid and Huber, Christian and Shukla, Khemraj and Karniadakis, George Em},
  journal={Journal of Geophysical Research: Solid Earth},
  volume={127},
  number={5},
  pages={e2021JB023120},
  year={2022},
  publisher={Wiley Online Library}
}

@article{rudenko2025reconstruction,
  title={Reconstruction of the mean velocity, pressure, and turbulent viscosity fields from background-oriented schlieren measurements using a physics-informed neural network},
  author={Rudenko, Yulia K and Vinnichenko, Nikolay A and Pushtaev, Alexei V and Plaksina, Yulia Yu and Uvarov, Alexander V},
  journal={Physics of Fluids},
  volume={37},
  number={9},
  pages={095106},
  year={2025},
  publisher={AIP Publishing}
}

@article{steinfurth2024assimilating,
  title={Assimilating experimental data of a mean three-dimensional separated flow using physics-informed neural networks},
  author={Steinfurth, B and Weiss, J},
  journal={Physics of Fluids},
  volume={36},
  number={1},
  pages={015131},
  year={2024},
  publisher={AIP Publishing}
}

\end{document}